\documentclass[fleqn,usenatbib]{mnras}
\usepackage{graphicx}
\usepackage{url}
\usepackage{longtable}
\usepackage{float}
\usepackage{subfigure}
\usepackage{newtxtext,newtxmath}
\usepackage[T1]{fontenc}
\usepackage{amsmath}
\usepackage{soul}
\usepackage{lscape}

\DeclareRobustCommand{\VAN}[3]{#2} \let\VANthebibliography\thebibliography
\def\thebibliography{\DeclareRobustCommand{\VAN}[3]{##3} \VANthebibliography}

\title[Weather Forecast of the Milky Way]{Weather Forecast of the Milky Way: Shear and Stellar feedback determine the lives of Galactic-scale filaments}

\author[Li, Zhou \& Chen]{
Guang-Xing Li,$^{1}$\thanks{gxli@ynu.edu.cn} 
Ji-Xuan Zhou,$^{1}$ 
Bing-Qiu Chen$^{1}$\thanks{bchen@ynu.edu.cn} \\
$^{1}$ South-Western Institute for Astronomy Research, Yunnan University, Chenggong District, Kunming 650091, P.\,R. China}

\begin{document}
\label{firstpage}

\maketitle

\begin{abstract}
    The interstellar medium (ISM) is an inseparable part of the Milky Way ecosystem
    whose evolutionary history remains a challenging question. We trace the evolution of the molecular ISM using a sample of Young Stellar Objects (YSO) association --molecular cloud complex (YSO-MC complex). We derive their three-dimensional (3D) velocities by combining the Gaia astrometric measurements of the YSO associations and the CO observations of the associated molecular clouds. Based on the 3D velocities, we simulate the motions of the YSO-MC complexes in the Galactic potential and forecast the ISM evolution by tracing the motions of the individual complexes, and reveal the roles of shear and stellar feedback in determining ISM evolution: Galactic shear stretches Galactic-scale molecular cloud complexes, such as the G120 Complex, into Galactic-scale filaments, and it also contributes to the destruction of the filaments; 
    while stellar feedback creates interconnected
    superbubbles whose expansion injects peculiar velocities into the ISM. 
    The Galactic-scale molecular gas clumps are often precursors of the filaments and the
    Galactic-scale filaments are transient structures under a constant stretch by shear. This evolutionary sequence sets a foundation to interpret other gas structures. 
\end{abstract} 



\section{Introduction}
The interstellar medium (ISM) is an inseparable part of the Milky Way ecosystem
whose evolutionary history remains a challenging question.   Studies find that
filaments of different sizes prevail in the molecular interstellar medium
\citep{2015ApJ...815...23Z,2016A&A...591A...5L,2016ApJS..226....9W}. Among them,
filaments whose sizes are larger than the thickness of the Milky Way molecular gas
disk are basic units of the molecular gases in the Milky Way and other similar disk
galaxies
\citep{Li2013,2014A&A...568A..73R,Goodman2014,Wang2015,2020Natur.578..237A}. These have been reproduced in global simulations of the Milky Way disk
\citep[e.g.][]{2000ApJ...540..797W,2006MNRAS.371.1663D,2015MNRAS.447.3390D,2016MNRAS.455.3640S,2020MNRAS.492.1594S,2022arXiv220503218L}.
The physics controlling the ISM evolution on the kpc scale and the nature of the
Galactic-scale filaments are still open questions. 

Young Stellar Objects (YSO) are newly-born stars and they inherit the motion of the gas where they originate from. The Gaia satellite \citep{2018a&a...616a...1g} allows for the measurement of motions of the YSOs on this sky plane. This, combined with radial velocity measurements, allow us to derive the 3D velocities of YSOs. In this paper, we trace the evolution of the molecular ISM using a sample of Young Stellar Objects (YSO) association \citep{2021arXiv211011595Z}--molecular cloud complex (YSO-MC complex). 
We derive their three-dimensional (3D) velocities by combining the Gaia astrometric measurements \citep{2018a&a...616a...1g} of the YSO associations and the CO observations of the associated molecular clouds \citep{2001ApJ...547..792D}. Based on the 3D velocities, we simulate the motions of the YSO-MC complexes in the Galactic potential and forecast the ISM evolution, where the clouds are assumed to be isolated objects, and effects from external gravitational potential \citep{2009MNRAS.393.1563B}, spiral density wave, are neglected.  This allows us to reveal the roles of shear and stellar feedback in determining ISM evolution.


\section{Data \& Method}

We start with a sample of Young Stellar Object (YSO) associations identified in
a previous paper \citep{2021arXiv211011595Z}. We identify the molecular cloud
counterparts of the YSO associations from the CO observations
\citep{2001ApJ...547..792D}. The entity, which contains a YSO association and
the associated molecular cloud, is noted as a YSO association--molecular cloud
complex (YSO-MC complex) in the current work. By combining the radial velocities
of the associated molecular clouds derived using CO lines and the transverse velocities of the YSO
associations, we obtain three-dimensional (3D) motions of a sample of YSO-MC
complexes. In this work, we adopt the convention that the Galaxy is viewed from
the north, such that the disk rotates clockwise. The Sun is 8.34 kpc from the
Galactic centner, and we are 0.0208 kpc above the Galactic disk midplane
\citep{Reid2014,2019MNRAS.482.1417B}. We first derive the velocities in the
Barycentric frame, then convert them to other reference frames (see online
supplementary materials).
\section{Results}
\subsection{Interpreting 3D motions}

\begin{figure*}
    \includegraphics[width = 0.9 \textwidth]{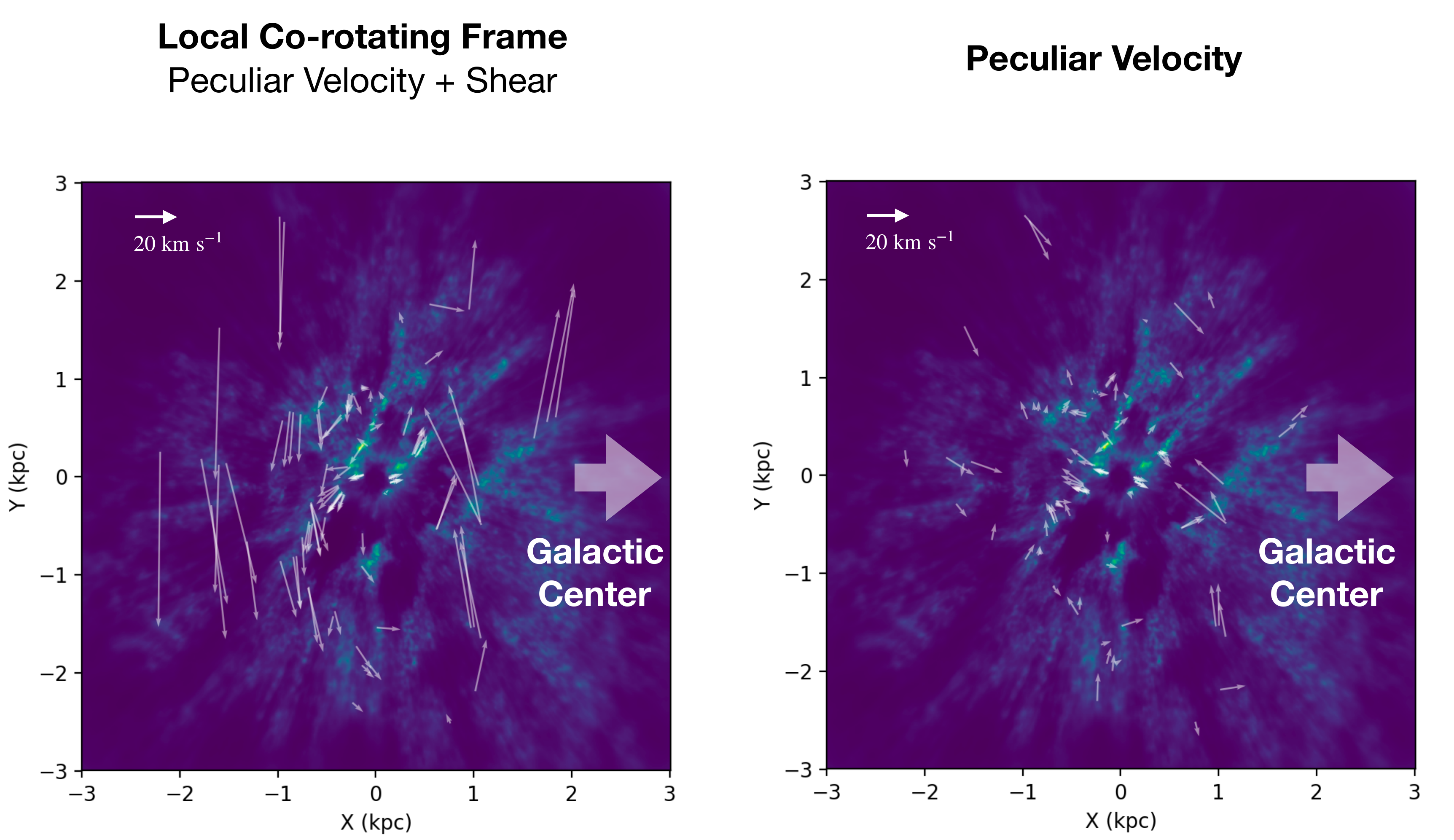}
    \caption{{\bf 3D velocities of our sample YSO-MC complexes.} {\bf Left:} Velocity field in the Local Co-rotating Frame (LCF). This frame is similar to the reference frame used in e.g. shearing-box simulations of disks, where the effect of shear on the gas is obvious. {\bf Right:} Map of the peculiar velocities.
    The background color-scale images are maps of interstellar gases traced by the dust \citep{2019A&A...625A.135L}.  \label{fig:velo} }
\end{figure*}

Shear occurs when gas located at different radii from the Galaxy center and
rotates at different angular speeds. In Fig.~\ref{fig:velo}, the effects of
shear appear as the radial dependence of the $Y$ direction velocities (the
dependence of $v_y$ on $X$). In the Solar neighborhood, the shear rate is
$\kappa = 2 A$ where $A$ is the Oort constant \citep{1927BAN.....3..275O} with
updated values \citep{2021MNRAS.504..199W}. We further define a reference frame
called the Local Co-Rotating Frame (LCF). The LCF follows a circular orbit, and
it is rotating such that its $x$-axis is locked to the center of the Galaxy.
Because of the rotation, the LCF is non-inertial, and gas in this frame
experiences the Coriolis force. We choose this frame as it allows us to observe
the effect of shear on the gas with convenience. 

 The clouds also contain peculiar velocities, which are the additional
 velocities from the cloud when measured against the mean motion of stars. To derive
 the peculiar velocities $v_{\rm peculiar}$ of our sample YSO-MC complexes in the LCF, we further subtract their shear component (see online supplementary materials). 
 The results are also shown in Fig. \ref{fig:velo}. Most
 sample YSO-MC complexes have $v_{\rm peculiar}$ of about 6\,km\,s$^{-1}$, while some of those sources exhibit very significant deviations. 

\subsection{Bubble-Induced Peculiar Velocities}

\begin{figure*}
    \includegraphics[width = 0.9 \textwidth]{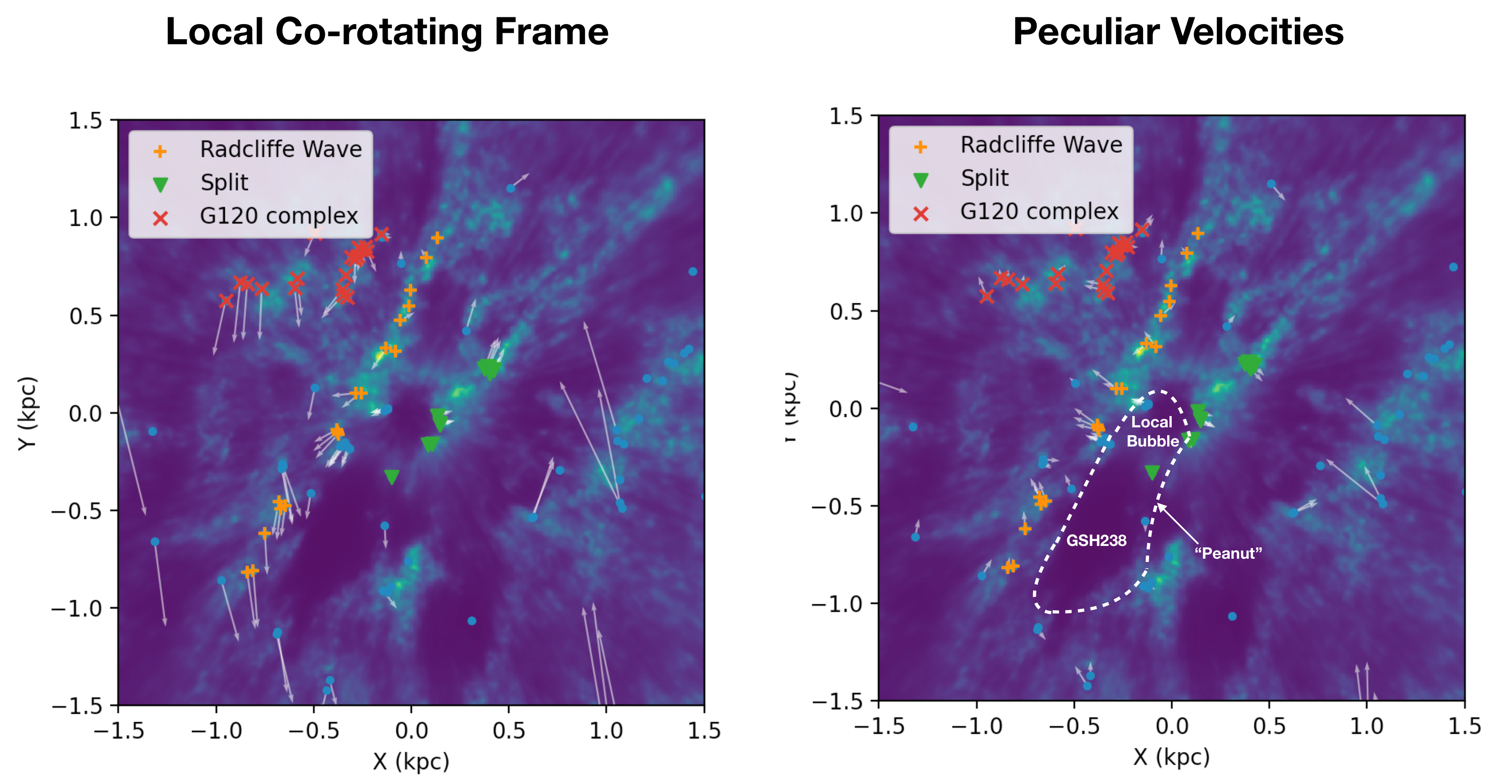}
    \caption{{\bf Velocities of our cataloged YSO-MC complexes around the Local
    Bubble and the GSH238 Bubble.} {\bf Left:} Velocity field in the Local
    Co-rotating Frame (LCF). This frame is similar to the reference frame used
    in e.g. shearing-box simulations of the accretion disk. {\bf Right:} Map of
    the peculiar 
    velocities. The background color-scale images are maps of
    interstellar gases traced by dust \citep{2019A&A...625A.135L}. Different
    symbols represent YSO-MC complexes belonging to different structures.  The
    Local Bubble, the GSH238 Bubble, and the boundary of the Peanut Superbubble
    are indicated. \label{fig:bubble} }
\end{figure*}

Superbubbles are cavities filled with hot gas and radiation. They are prominent
structures resulting from energies produced during the lifetime of stars,
including energy from the stellar wind, radiation from massive stars, and the
supernova explosions \citep{1988ARA&A..26..145T}. The Sun is located inside one
such bubble called the Local Bubble \citep{1987ARA&A..25..303C}. The Local
Bubble appears to be connected to a neighboring bubble called the GSH238 Bubble
\citep{1998ApJ...498..689H}, and from a dust
map\citep{2019A&A...625A.135L,2019MNRAS.483.4277C}, these two bubbles form a
peanut-shaped cavity which we call the ``Peanut Superbubble"
(Fig.~\ref{fig:bubble}). Systems like this should be common in our Galaxy, as
previous studies modeled the evolution of superbubble populations in Milky
Way-like galaxy disks and concluded that bubble mergers are inevitable
\citep{2015A&A...578A.113K}. An expansion-induced peculiar velocity of  5.5 $\rm
km\;s^{-1}$ can be measured from the map of the peculiar velocities at a scale
of 0.3  kpc. This expansion velocity is small compared to other bubbles,
indicating that the expansion of the Local Bubble has reached its end
\citep{2015A&A...578A.113K}. The expansion rate, which is around 30 $\rm km \;
s^{-1}\; kpc^{-1}$, is significant yet comparable to that from shear measured in
terms of Oort constant   \citep{1927BAN.....3..275O}, where  the shear rate is
$\kappa = 2 \; A \approx  30\; \rm km \; s^{-1}\; kpc^{-1} $. 

 \subsection{Forecasting ISM evolution}

 With the measurements of the 3D positions and 3D velocities of our sample
 YSO-MC complexes, we can forecast the future evolution of the system of clouds
 by tracking the motions of the individual YSO-MC complexes in the
 Galactic-scale potential (see online supplementary materials). We integrate the
 system to a time of 30 Myr, which is comparable to the shear time.

 Our approach should capture the crucial causal connection between the current
 motion of a cloud and its future positions. There are some other physical
 processes, which are not modeled but should be negligible for our purposes.
 These include the stellar feedback processes and the self-gravity from the gas.
 In galaxies, multiple supernovas can act collectively, creating superbubbles
 \citep{1988ARA&A..26..145T}. They can reach the size of $\sim 10^2$\,pc, and
 affect the dynamics of the ISM during their expansions. In our case, the
 expansion of the Local Bubble and the GSH238 Bubble leads to a peculiar
 velocity of about 6\,$\rm km\,s^{-1}$. This small value of peculiar velocity is
 typical for bubbles whose expansion has reached an end 
 \citep{2015A&A...578A.113K}. We thus do not expect additional velocity
 injections from existing bubbles. 

Self-gravity from gas might give rise to some additional alignments. We conclude
that the self-gravity of the molecular gas is not important on the Galactic
scale. To evaluate its importance, we focus on the dense regions, i.e. the
Galactic-scale filaments. The line-mass of the Galactic-scale filaments has
$\delta_{\rm ml} \lesssim 1000 M_{\odot}\;\rm pc$ \citep{Li2016}. The
characteristic velocity signifying the importance of the gas self-gravity can be
estimated by $v_{\rm c} \approx \sqrt{G \delta_{\rm ml} } \approx 2\;\rm km\;
s^{-1}$, which is small
compared to the shear effect. 

\begin{figure*}
    \includegraphics[width = 0.49\textwidth]{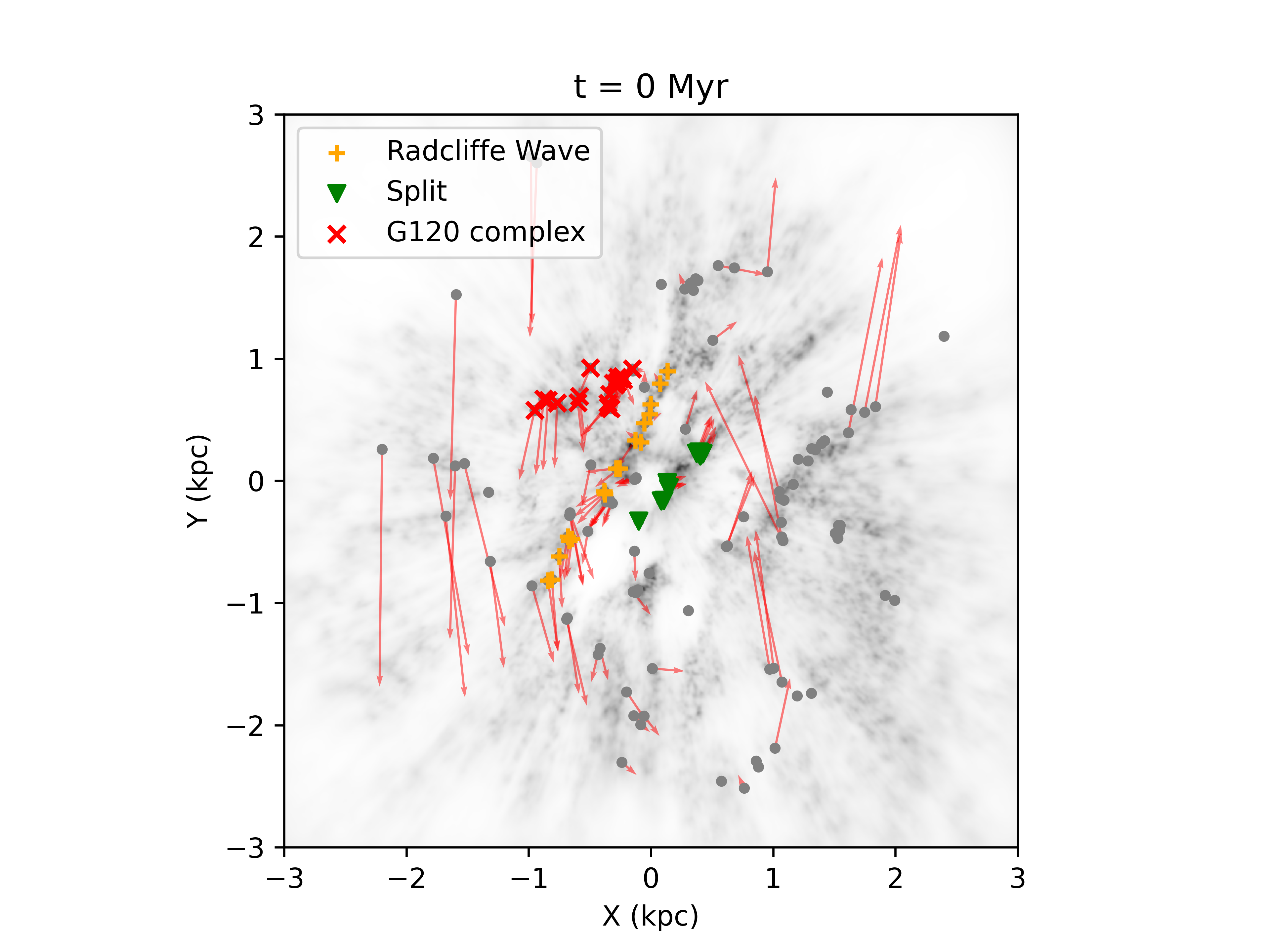}
    \includegraphics[width = 0.49\textwidth]{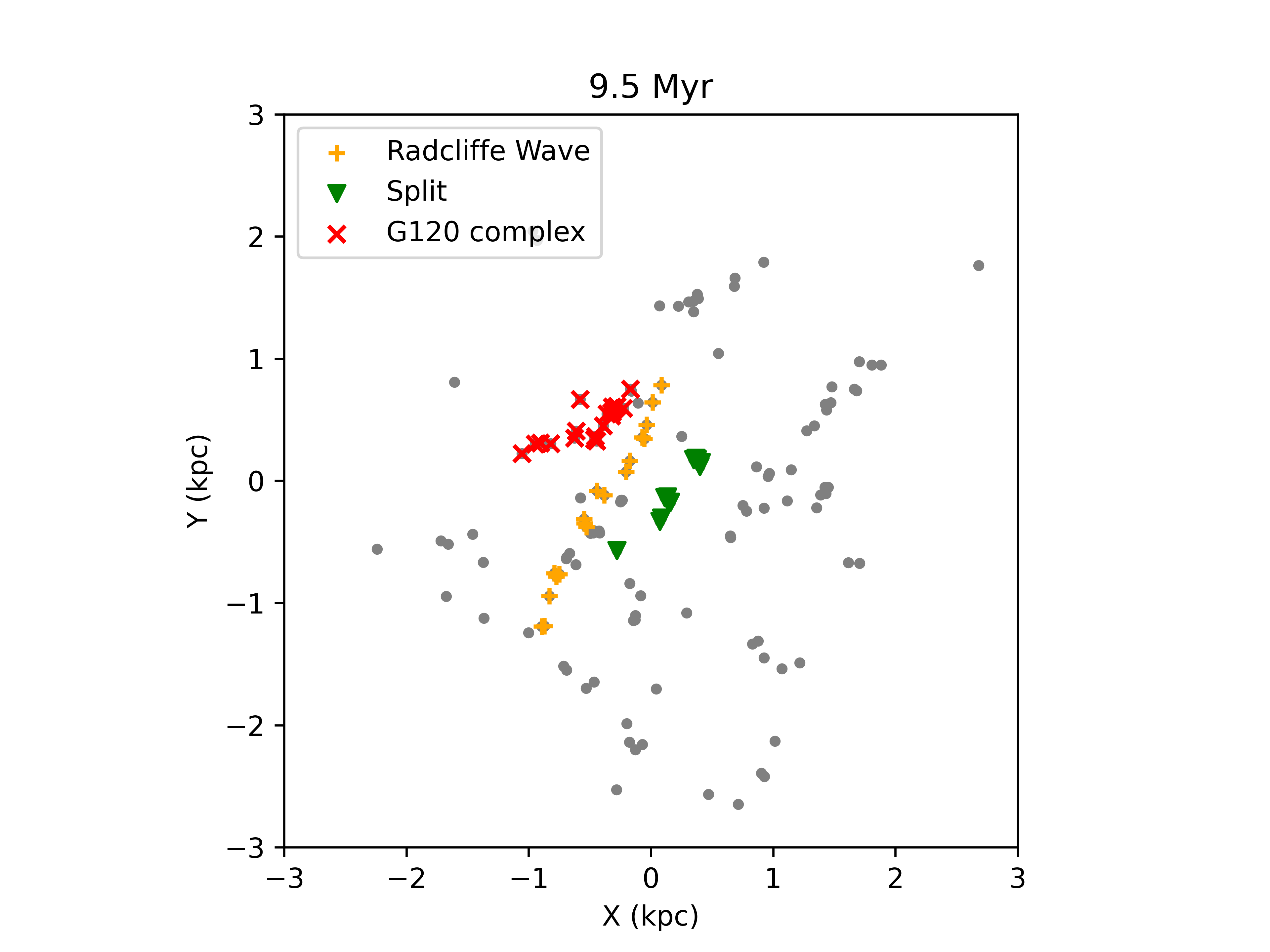} \\
    \includegraphics[width = 0.49\textwidth]{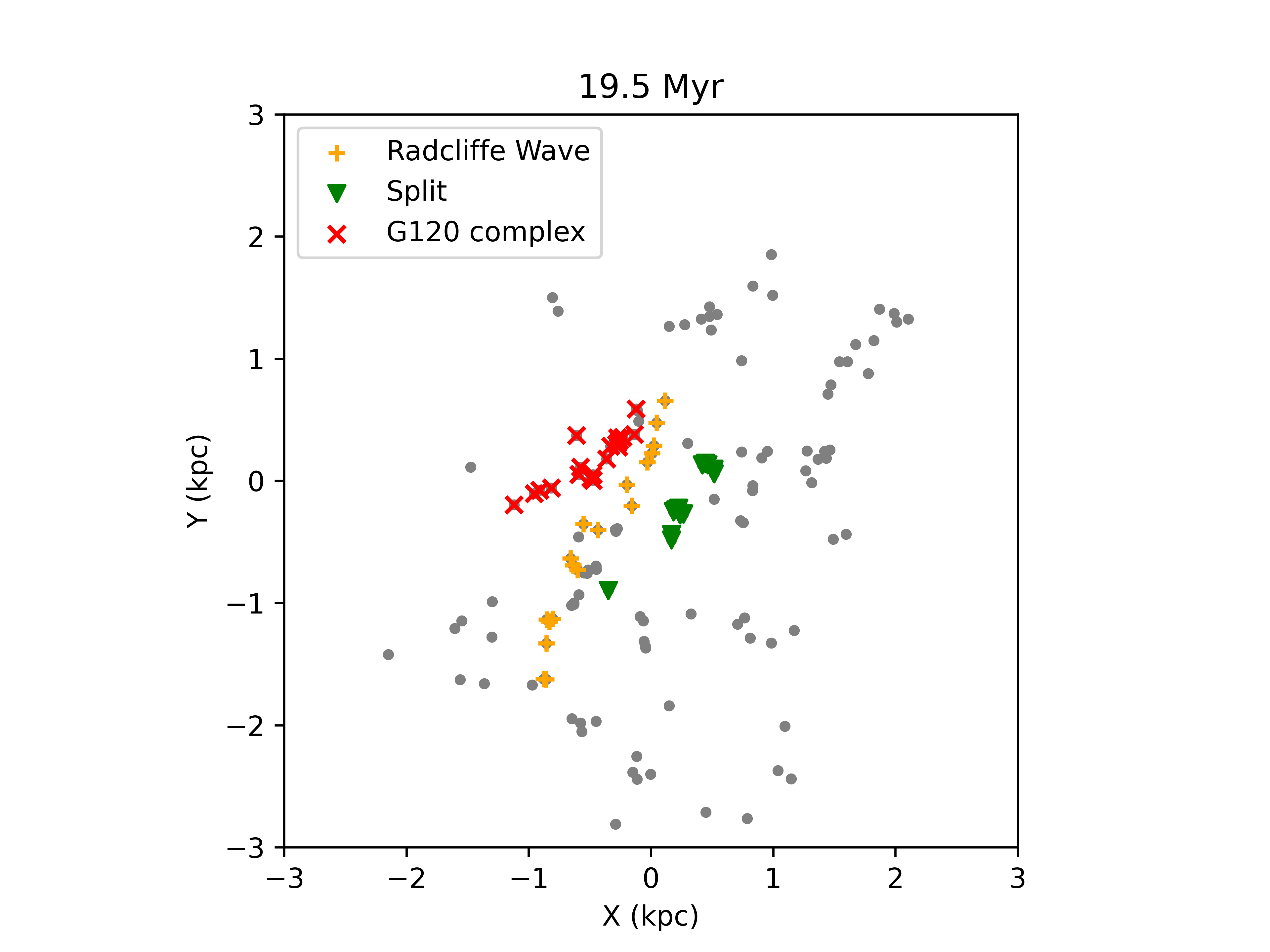} 
    \includegraphics[width = 0.49\textwidth]{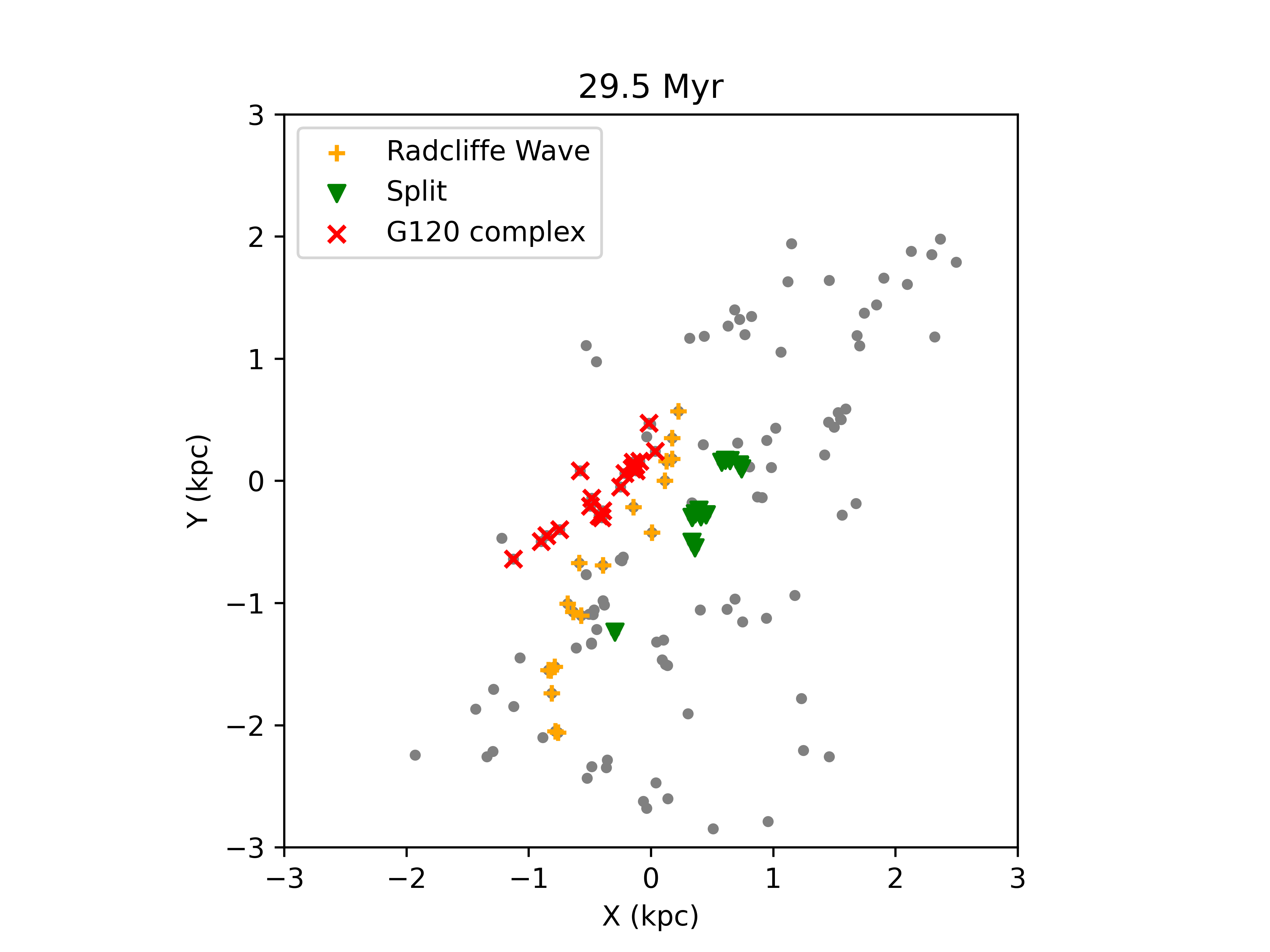}

\caption{{\bf Forecasting the evolution of molecular ISM.}  Here, different structures are indicated by different symbols. We forecast the ISM evolution by modeling the motions of the individual YSO-MC complexes in the Galactic potential.
At $t\approx 30 \;\rm Myr$, The ``split'' and the Radcliffe Wave will be stretched significantly, while the G120 Complex can evolve into a long filament of width $L \approx \rm 1\; kpc$.  In the first panel, we added a color-scale image of the gas distribution traced by dust, and the 3D velocities measured in the LCF for reference. 
\label{fig:movie}    }
\end{figure*}

Another assumption adopted in the current work is that the clouds needs to be long-lived, with $t_{\rm cloud} > t_{\rm simulation} =t_{\rm shear}$. This condition can be satisfied, as the molecular clouds are
 estimated to live up to a few tens of Myr \citep{2015ApJ...806...72M,2021RNAAS...5..222K,2021arXiv211011595Z}, which is comparable or longer than
 the shear time \footnote{This, of course, dependents on the lifetime of cloud, and the forecast becomes inaccurate if clouds are short \citep{2001ApJ...562..852H}. Recently studies point to a picture where gas in a cloud can stay for a long time in the cold phase and circulate between different spiral arms \citep{2001MNRAS.327..663P,2013MNRAS.432..653D}.  }.
 Our evolution forecast results are plotted in Fig.~\ref{fig:movie}, and
 a movie is attached. 

\subsection{Birth and death of Galactic-scale Filaments}

In this paper, the term ``Galactic-scale filaments" \citep{Li2013,2014A&A...568A..73R,Goodman2014,Wang2015}, refers to filaments whose length exceeds the thickness of the Galactic molecular gas disk ($\sim$ 100\,pc \citep{2015ARA&A..53..583H,Guo2021}). In the Solar vicinity, this includes the Radcliffe Wave \citep{2020Natur.578..237A,2022arXiv220503218L}, and the ``Split'' \citep{2019A&A...625A.135L,2020MNRAS.493..351C}. One end of the  ``Split'' stays at the edge of the Peanut Superbubble, which is a
a connected entity made of the Local Bubble and the GH238 Bubble \citep{1998ApJ...498..689H}.  

This picture of a dynamically-evolving Galaxy molecular gas, first revealed in
this study, bears a striking resemblance to the results from simulations of disk
Galaxies \citep{2011MNRAS.413.2935D,2017MNRAS.470.4261D}. We trace the evolution
of three structures: the G120 Complex, the Radcliffe Wave, and the ``Split''. We
propose that the three structures form an evolutionary sequence, with the G120
Complex being a filament precursor, the Radcliffe Wave being a typical example
of the Galactic-scale filament and the ``Split'' being a Galactic-scale filament
at a slightly later stage of evolution. Our conclusion is based on the evolution
forecast results: As the simulation goes, the G120 Complex is turned into a
filament, and the Radcliffe Wave, as well as the ``Split'', is stretched
significantly, where the ``Split'' would be barely recognizable by then. Here,
the disruption is mostly caused by shear alone. New supernova explosions should
also contribute to the disruption of the filament,  further shortening their
lifetimes.

\begin{figure*}
    \includegraphics[width = 0.7\textwidth]{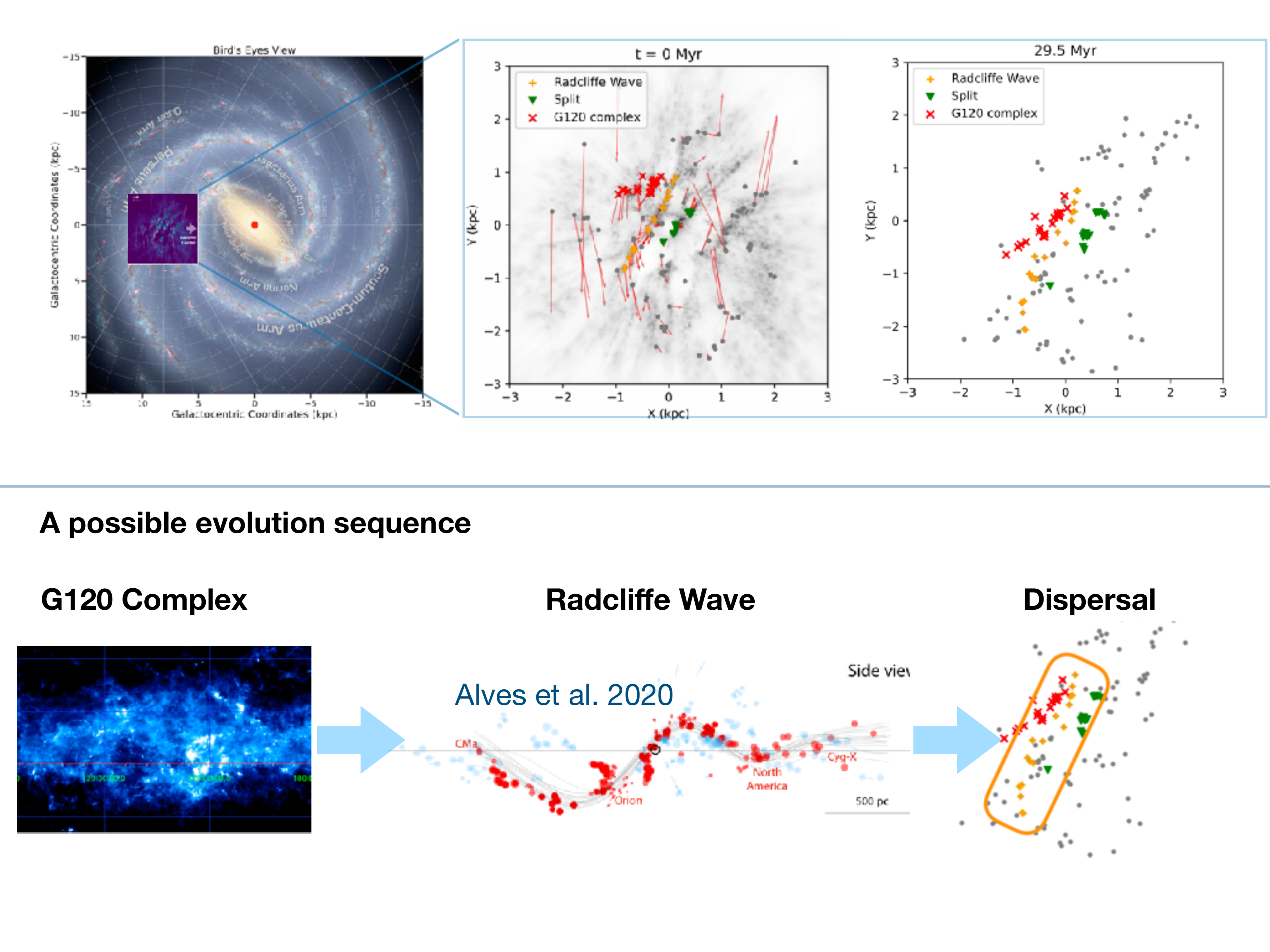}
\caption{{\bf Evolution sequence of Galactic scale structures.} The map of the G120 complex is taken at 870 $\mu$m by the Planck Satellite \citep{2014planck}. A illustrate of the 3D structure of the Radcliffe Wave is taken from \citet{2020Natur.578..237A}.  \label{fig:one}}
\end{figure*}

 \section{Concluding remarks}

 We present the first study of the evolution of the local molecular ISM by combining the proper motions and parallaxes of the YSO associations with radial velocity measurements towards their molecular cloud counterparts. The complete kinematic information reveals a vivid picture of dynamically-evolving Galactic molecular gas.
 Through an approach shall be called the
 Milky Way Weather Forecast, we predict the future configuration of the
 molecular ISM by modeling the movement of individual clouds in the Galactic potential. This allows us to identify an evolutionary sequence, where
 giant gas clumps get stretched into Galactic-scale filaments, which disperse
 further. The G120 Complex is one filament precursor, which deserves further investigations. 

 We conclude that hidden under a complex appearance, it is the shear and stellar
 feedback which control the evolution of the molecular ISM on the kpc scale. In
 addition, we identify a sequence where Galactic-scale gas clumps evolve into
 filaments that disperse because of shear and stellar feedback. The filaments
 are thus transient objects with lifetimes comparable to the shear time. During
 their lifetimes, the expansions of superbubbles can lead to local deformation
 and injections of peculiar velocities. This dynamically evolving Galactic
 molecular gas revealed in this paper sets a solid foundation to link different
 observations and discover new patterns.\\

 Additional plots and a movie is available at \url{https://gxli.github.io/ISM-6D/movie.html}.

  \section*{Acknowledgements}
  We thank the referee for a careful reading of the paper and for the constructive comments. 
GXL acknowledges supports from National
Natural Science Foundation of China grant W820301904 and 12033005. BQC is
supported by the National Key R\&D Program of China No. 2019YFA0405500, National
Natural Science Foundation of China 12173034, 11803029 and 11833006, and the
science research grants from the China Manned Space Project with NO.
CMS-CSST-2021- A09, CMS-CSST-2021-A08 and CMS-CSST-2021-B03. This work presents
results from the European Space Agency (ESA) space mission Gaia. Gaia data are
being processed by the Gaia Data Processing and Analysis Consortium (DPAC).
Funding for the DPAC is provided by national institutions, in particular the
institutions participating in the Gaia MultiLateral Agreement (MLA). The Gaia
mission website is \url{https://www.cosmos.esa.int/gaia}. The Gaia archive
website is \url{https://archives.esac.esa.int/gaia}. This research has used
Astropy and Galpy.


\section*{DATA AVAILABILITY}
The paper makes use of published data from \citet{2016An} and Gaia DR2 \citep{GAIADR2}. Our Table containing the location and classification type information will be available as Zhou, Li \& Chen 2022 in prep.

\bibliographystyle{mnras}
\bibliography{paper.bib}

\begin{thebibliography}{}
\makeatletter
\relax
\def\mn@urlcharsother{\let\do\@makeother \do\$\do\&\do\#\do\^\do\_\do\%\do\~}
\def\mn@doi{\begingroup\mn@urlcharsother \@ifnextchar [ {\mn@doi@}
  {\mn@doi@[]}}
\def\mn@doi@[#1]#2{\def\@tempa{#1}\ifx\@tempa\@empty \href
  {http://dx.doi.org/#2} {doi:#2}\else \href {http://dx.doi.org/#2} {#1}\fi
  \endgroup}
\def\mn@eprint#1#2{\mn@eprint@#1:#2::\@nil}
\def\mn@eprint@arXiv#1{\href {http://arxiv.org/abs/#1} {{\tt arXiv:#1}}}
\def\mn@eprint@dblp#1{\href {http://dblp.uni-trier.de/rec/bibtex/#1.xml}
  {dblp:#1}}
\def\mn@eprint@#1:#2:#3:#4\@nil{\def\@tempa {#1}\def\@tempb {#2}\def\@tempc
  {#3}\ifx \@tempc \@empty \let \@tempc \@tempb \let \@tempb \@tempa \fi \ifx
  \@tempb \@empty \def\@tempb {arXiv}\fi \@ifundefined
  {mn@eprint@\@tempb}{\@tempb:\@tempc}{\expandafter \expandafter \csname
  mn@eprint@\@tempb\endcsname \expandafter{\@tempc}}}

\bibitem[\protect\citeauthoryear{{Alves} et~al.,}{{Alves}
  et~al.}{2020}]{2020Natur.578..237A}
{Alves} J.,  et~al., 2020, \mn@doi [\nat] {10.1038/s41586-019-1874-z}, \href
  {https://ui.adsabs.harvard.edu/abs/2020Natur.578..237A} {578, 237}

\bibitem[\protect\citeauthoryear{{Ballesteros-Paredes}, {G{\'o}mez}, {Pichardo}
   \& {V{\'a}zquez-Semadeni}}{{Ballesteros-Paredes}
  et~al.}{2009}]{2009MNRAS.393.1563B}
{Ballesteros-Paredes} J.,  {G{\'o}mez} G.~C.,  {Pichardo} B.,
  {V{\'a}zquez-Semadeni} E.,  2009, \mn@doi [\mnras]
  {10.1111/j.1365-2966.2008.14278.x}, \href
  {https://ui.adsabs.harvard.edu/abs/2009MNRAS.393.1563B} {393, 1563}

\bibitem[\protect\citeauthoryear{{Bennett} \& {Bovy}}{{Bennett} \&
  {Bovy}}{2019}]{2019MNRAS.482.1417B}
{Bennett} M.,  {Bovy} J.,  2019, \mn@doi [\mnras] {10.1093/mnras/sty2813},
  \href {https://ui.adsabs.harvard.edu/abs/2019MNRAS.482.1417B} {482, 1417}

\bibitem[\protect\citeauthoryear{{Chen} et~al.,}{{Chen}
  et~al.}{2019}]{2019MNRAS.483.4277C}
{Chen} B.~Q.,  et~al., 2019, \mn@doi [\mnras] {10.1093/mnras/sty3341}, \href
  {https://ui.adsabs.harvard.edu/abs/2019MNRAS.483.4277C} {483, 4277}

\bibitem[\protect\citeauthoryear{{Chen} et~al.,}{{Chen}
  et~al.}{2020}]{2020MNRAS.493..351C}
{Chen} B.~Q.,  et~al., 2020, \mn@doi [\mnras] {10.1093/mnras/staa235}, \href
  {https://ui.adsabs.harvard.edu/abs/2020MNRAS.493..351C} {493, 351}

\bibitem[\protect\citeauthoryear{{Cox} \& {Reynolds}}{{Cox} \&
  {Reynolds}}{1987}]{1987ARA&A..25..303C}
{Cox} D.~P.,  {Reynolds} R.~J.,  1987, \mn@doi [\araa]
  {10.1146/annurev.aa.25.090187.001511}, \href
  {https://ui.adsabs.harvard.edu/abs/1987ARA&A..25..303C} {25, 303}

\bibitem[\protect\citeauthoryear{{Dame}, {Hartmann}  \& {Thaddeus}}{{Dame}
  et~al.}{2001}]{2001ApJ...547..792D}
{Dame} T.~M.,  {Hartmann} D.,   {Thaddeus} P.,  2001, \mn@doi [\apj]
  {10.1086/318388}, \href
  {https://ui.adsabs.harvard.edu/abs/2001ApJ...547..792D} {547, 792}

\bibitem[\protect\citeauthoryear{{Dobbs}}{{Dobbs}}{2015}]{2015MNRAS.447.3390D}
{Dobbs} C.~L.,  2015, \mn@doi [\mnras] {10.1093/mnras/stu2585}, \href
  {https://ui.adsabs.harvard.edu/abs/2015MNRAS.447.3390D} {447, 3390}

\bibitem[\protect\citeauthoryear{{Dobbs} \& {Pringle}}{{Dobbs} \&
  {Pringle}}{2013}]{2013MNRAS.432..653D}
{Dobbs} C.~L.,  {Pringle} J.~E.,  2013, \mn@doi [\mnras]
  {10.1093/mnras/stt508}, \href
  {https://ui.adsabs.harvard.edu/abs/2013MNRAS.432..653D} {432, 653}

\bibitem[\protect\citeauthoryear{{Dobbs}, {Bonnell}  \& {Pringle}}{{Dobbs}
  et~al.}{2006}]{2006MNRAS.371.1663D}
{Dobbs} C.~L.,  {Bonnell} I.~A.,   {Pringle} J.~E.,  2006, \mn@doi [\mnras]
  {10.1111/j.1365-2966.2006.10794.x}, \href
  {http://adsabs.harvard.edu/abs/2006MNRAS.371.1663D} {371, 1663}

\bibitem[\protect\citeauthoryear{{Dobbs}, {Burkert}  \& {Pringle}}{{Dobbs}
  et~al.}{2011}]{2011MNRAS.413.2935D}
{Dobbs} C.~L.,  {Burkert} A.,   {Pringle} J.~E.,  2011, \mn@doi [\mnras]
  {10.1111/j.1365-2966.2011.18371.x}, \href
  {http://adsabs.harvard.edu/abs/2011MNRAS.413.2935D} {413, 2935}

\bibitem[\protect\citeauthoryear{{Duarte-Cabral} \& {Dobbs}}{{Duarte-Cabral} \&
  {Dobbs}}{2017}]{2017MNRAS.470.4261D}
{Duarte-Cabral} A.,  {Dobbs} C.~L.,  2017, \mn@doi [\mnras]
  {10.1093/mnras/stx1524}, \href
  {https://ui.adsabs.harvard.edu/abs/2017MNRAS.470.4261D} {470, 4261}

\bibitem[\protect\citeauthoryear{{Gaia Collaboration} et~al.,}{{Gaia
  Collaboration} et~al.}{2018a}]{2018a&a...616a...1g}
{Gaia Collaboration} et~al., 2018a, \mn@doi [\aap]
  {10.1051/0004-6361/201833051}, \href
  {https://ui.adsabs.harvard.edu/abs/2018A&A...616A...1G} {616, A1}

\bibitem[\protect\citeauthoryear{{Gaia Collaboration} et~al.,}{{Gaia
  Collaboration} et~al.}{2018b}]{GAIADR2}
{Gaia Collaboration} et~al., 2018b, \mn@doi [\aap]
  {10.1051/0004-6361/201833051}, \href
  {https://ui.adsabs.harvard.edu/abs/2018A&A...616A...1G} {616, A1}

\bibitem[\protect\citeauthoryear{{Goodman} et~al.,}{{Goodman}
  et~al.}{2014}]{Goodman2014}
{Goodman} A.~A.,  et~al., 2014, \mn@doi [\apj] {10.1088/0004-637X/797/1/53},
  \href {http://adsabs.harvard.edu/abs/2014ApJ...797...53G} {797, 53}

\bibitem[\protect\citeauthoryear{{Guo} et~al.,}{{Guo} et~al.}{2021}]{Guo2021}
{Guo} H.~L.,  et~al., 2021, \mn@doi [\apj] {10.3847/1538-4357/abc68a}, \href
  {https://ui.adsabs.harvard.edu/abs/2021ApJ...906...47G} {906, 47}

\bibitem[\protect\citeauthoryear{{Hartmann}, {Ballesteros-Paredes}  \&
  {Bergin}}{{Hartmann} et~al.}{2001}]{2001ApJ...562..852H}
{Hartmann} L.,  {Ballesteros-Paredes} J.,   {Bergin} E.~A.,  2001, \mn@doi
  [\apj] {10.1086/323863}, \href
  {https://ui.adsabs.harvard.edu/abs/2001ApJ...562..852H} {562, 852}

\bibitem[\protect\citeauthoryear{{Heiles}}{{Heiles}}{1998}]{1998ApJ...498..689H}
{Heiles} C.,  1998, \mn@doi [\apj] {10.1086/305574}, \href
  {https://ui.adsabs.harvard.edu/abs/1998ApJ...498..689H} {498, 689}

\bibitem[\protect\citeauthoryear{{Heyer} \& {Dame}}{{Heyer} \&
  {Dame}}{2015}]{2015ARA&A..53..583H}
{Heyer} M.,  {Dame} T.~M.,  2015, \mn@doi [\araa]
  {10.1146/annurev-astro-082214-122324}, \href
  {https://ui.adsabs.harvard.edu/abs/2015ARA&A..53..583H} {53, 583}

\bibitem[\protect\citeauthoryear{{Koda}}{{Koda}}{2021}]{2021RNAAS...5..222K}
{Koda} J.,  2021, \mn@doi [Research Notes of the American Astronomical Society]
  {10.3847/2515-5172/ac2d34}, \href
  {https://ui.adsabs.harvard.edu/abs/2021RNAAS...5..222K} {5, 222}

\bibitem[\protect\citeauthoryear{{Krause} et~al.,}{{Krause}
  et~al.}{2015}]{2015A&A...578A.113K}
{Krause} M. G.~H.,  et~al., 2015, \mn@doi [\aap] {10.1051/0004-6361/201525847},
  \href {https://ui.adsabs.harvard.edu/abs/2015A&A...578A.113K} {578, A113}

\bibitem[\protect\citeauthoryear{{Lallement}, {Babusiaux}, {Vergely}, {Katz},
  {Arenou}, {Valette}, {Hottier}  \& {Capitanio}}{{Lallement}
  et~al.}{2019}]{2019A&A...625A.135L}
{Lallement} R.,  {Babusiaux} C.,  {Vergely} J.~L.,  {Katz} D.,  {Arenou} F.,
  {Valette} B.,  {Hottier} C.,   {Capitanio} L.,  2019, \mn@doi [\aap]
  {10.1051/0004-6361/201834695}, \href
  {https://ui.adsabs.harvard.edu/abs/2019A&A...625A.135L} {625, A135}

\bibitem[\protect\citeauthoryear{{Li} \& {Chen}}{{Li} \&
  {Chen}}{2022}]{2022arXiv220503218L}
{Li} G.-X.,  {Chen} B.-Q.,  2022, arXiv e-prints, \href
  {https://ui.adsabs.harvard.edu/abs/2022arXiv220503218L} {p. arXiv:2205.03218}

\bibitem[\protect\citeauthoryear{{Li}, {Wyrowski}, {Menten}  \&
  {Belloche}}{{Li} et~al.}{2013}]{Li2013}
{Li} G.-X.,  {Wyrowski} F.,  {Menten} K.,   {Belloche} A.,  2013, \mn@doi
  [\aap] {10.1051/0004-6361/201322411}, \href
  {http://adsabs.harvard.edu/abs/2013A%26A...559A..34L} {559, A34}

\bibitem[\protect\citeauthoryear{{Li}, {Urquhart}, {Leurini}, {Csengeri},
  {Wyrowski}, {Menten}  \& {Schuller}}{{Li}
  et~al.}{2016a}]{2016A&A...591A...5L}
{Li} G.-X.,  {Urquhart} J.~S.,  {Leurini} S.,  {Csengeri} T.,  {Wyrowski} F.,
  {Menten} K.~M.,   {Schuller} F.,  2016a, \mn@doi [\aap]
  {10.1051/0004-6361/201527468}, \href
  {https://ui.adsabs.harvard.edu/abs/2016A&A...591A...5L} {591, A5}

\bibitem[\protect\citeauthoryear{{Li}, {Urquhart}, {Leurini}, {Csengeri},
  {Wyrowski}, {Menten}  \& {Schuller}}{{Li} et~al.}{2016b}]{Li2016}
{Li} G.-X.,  {Urquhart} J.~S.,  {Leurini} S.,  {Csengeri} T.,  {Wyrowski} F.,
  {Menten} K.~M.,   {Schuller} F.,  2016b, \mn@doi [\aap]
  {10.1051/0004-6361/201527468}, \href
  {http://adsabs.harvard.edu/abs/2016A%26A...591A...5L} {591, A5}

\bibitem[\protect\citeauthoryear{{Marton}, {T{\'o}th}, {Paladini}, {Kun},
  {Zahorecz}, {McGehee}  \& {Kiss}}{{Marton} et~al.}{2016}]{2016An}
{Marton} G.,  {T{\'o}th} L.~V.,  {Paladini} R.,  {Kun} M.,  {Zahorecz} S.,
  {McGehee} P.,   {Kiss} C.,  2016, \mn@doi [\mnras] {10.1093/mnras/stw398},
  \href {https://ui.adsabs.harvard.edu/abs/2016MNRAS.458.3479M} {458, 3479}

\bibitem[\protect\citeauthoryear{{Meidt} et~al.,}{{Meidt}
  et~al.}{2015}]{2015ApJ...806...72M}
{Meidt} S.~E.,  et~al., 2015, \mn@doi [\apj] {10.1088/0004-637X/806/1/72},
  \href {https://ui.adsabs.harvard.edu/abs/2015ApJ...806...72M} {806, 72}

\bibitem[\protect\citeauthoryear{{Oort}}{{Oort}}{1927}]{1927BAN.....3..275O}
{Oort} J.~H.,  1927, \bain, \href
  {https://ui.adsabs.harvard.edu/abs/1927BAN.....3..275O} {3, 275}

\bibitem[\protect\citeauthoryear{{Planck Collaboration} et~al.,}{{Planck
  Collaboration} et~al.}{2014}]{2014planck}
{Planck Collaboration} et~al., 2014, \mn@doi [\aap]
  {10.1051/0004-6361/201321591}, \href
  {https://ui.adsabs.harvard.edu/abs/2014A&A...571A..16P} {571, A16}

\bibitem[\protect\citeauthoryear{{Pringle}, {Allen}  \& {Lubow}}{{Pringle}
  et~al.}{2001}]{2001MNRAS.327..663P}
{Pringle} J.~E.,  {Allen} R.~J.,   {Lubow} S.~H.,  2001, \mn@doi [\mnras]
  {10.1046/j.1365-8711.2001.04777.x}, \href
  {https://ui.adsabs.harvard.edu/abs/2001MNRAS.327..663P} {327, 663}

\bibitem[\protect\citeauthoryear{{Ragan}, {Henning}, {Tackenberg}, {Beuther},
  {Johnston}, {Kainulainen}  \& {Linz}}{{Ragan}
  et~al.}{2014}]{2014A&A...568A..73R}
{Ragan} S.~E.,  {Henning} T.,  {Tackenberg} J.,  {Beuther} H.,  {Johnston}
  K.~G.,  {Kainulainen} J.,   {Linz} H.,  2014, \mn@doi [\aap]
  {10.1051/0004-6361/201423401}, \href
  {http://adsabs.harvard.edu/abs/2014A%26A...568A..73R} {568, A73}

\bibitem[\protect\citeauthoryear{{Reid} et~al.,}{{Reid}
  et~al.}{2009}]{2009ApJ...700..137R}
{Reid} M.~J.,  et~al., 2009, \mn@doi [\apj] {10.1088/0004-637X/700/1/137},
  \href {https://ui.adsabs.harvard.edu/abs/2009ApJ...700..137R} {700, 137}

\bibitem[\protect\citeauthoryear{Reid et~al.,}{Reid et~al.}{2014}]{Reid2014}
Reid M.~J.,  et~al., 2014, \mn@doi [The Astrophysical Journal]
  {10.1088/0004-637x/783/2/130}, 783, 130

\bibitem[\protect\citeauthoryear{{Smith}, {Glover}, {Klessen}  \&
  {Fuller}}{{Smith} et~al.}{2016}]{2016MNRAS.455.3640S}
{Smith} R.~J.,  {Glover} S. C.~O.,  {Klessen} R.~S.,   {Fuller} G.~A.,  2016,
  \mn@doi [\mnras] {10.1093/mnras/stv2559}, \href
  {https://ui.adsabs.harvard.edu/abs/2016MNRAS.455.3640S} {455, 3640}

\bibitem[\protect\citeauthoryear{{Smith} et~al.,}{{Smith}
  et~al.}{2020}]{2020MNRAS.492.1594S}
{Smith} R.~J.,  et~al., 2020, \mn@doi [\mnras] {10.1093/mnras/stz3328}, \href
  {https://ui.adsabs.harvard.edu/abs/2020MNRAS.492.1594S} {492, 1594}

\bibitem[\protect\citeauthoryear{{Tenorio-Tagle} \&
  {Bodenheimer}}{{Tenorio-Tagle} \& {Bodenheimer}}{1988}]{1988ARA&A..26..145T}
{Tenorio-Tagle} G.,  {Bodenheimer} P.,  1988, \mn@doi [\araa]
  {10.1146/annurev.aa.26.090188.001045}, \href
  {https://ui.adsabs.harvard.edu/abs/1988ARA&A..26..145T} {26, 145}

\bibitem[\protect\citeauthoryear{{Wada}, {Spaans}  \& {Kim}}{{Wada}
  et~al.}{2000}]{2000ApJ...540..797W}
{Wada} K.,  {Spaans} M.,   {Kim} S.,  2000, \mn@doi [\apj] {10.1086/309347},
  \href {https://ui.adsabs.harvard.edu/abs/2000ApJ...540..797W} {540, 797}

\bibitem[\protect\citeauthoryear{{Wang}, {Testi}, {Ginsburg}, {Walmsley},
  {Molinari}  \& {Schisano}}{{Wang} et~al.}{2015}]{Wang2015}
{Wang} K.,  {Testi} L.,  {Ginsburg} A.,  {Walmsley} C.~M.,  {Molinari} S.,
  {Schisano} E.,  2015, \mn@doi [\mnras] {10.1093/mnras/stv735}, \href
  {http://adsabs.harvard.edu/abs/2015MNRAS.450.4043W} {450, 4043}

\bibitem[\protect\citeauthoryear{{Wang}, {Testi}, {Burkert}, {Walmsley},
  {Beuther}  \& {Henning}}{{Wang} et~al.}{2016}]{2016ApJS..226....9W}
{Wang} K.,  {Testi} L.,  {Burkert} A.,  {Walmsley} C.~M.,  {Beuther} H.,
  {Henning} T.,  2016, \mn@doi [\apjs] {10.3847/0067-0049/226/1/9}, \href
  {http://adsabs.harvard.edu/abs/2016ApJS..226....9W} {226, 9}

\bibitem[\protect\citeauthoryear{{Wang}, {Zhang}, {Huang}, {Chen}, {Wang}  \&
  {Wang}}{{Wang} et~al.}{2021}]{2021MNRAS.504..199W}
{Wang} F.,  {Zhang} H.~W.,  {Huang} Y.,  {Chen} B.~Q.,  {Wang} H.~F.,   {Wang}
  C.,  2021, \mn@doi [\mnras] {10.1093/mnras/stab848}, \href
  {https://ui.adsabs.harvard.edu/abs/2021MNRAS.504..199W} {504, 199}

\bibitem[\protect\citeauthoryear{{Zhou}, {Li}  \& {Chen}}{{Zhou}
  et~al.}{2021}]{2021arXiv211011595Z}
{Zhou} J.-X.,  {Li} G.-X.,   {Chen} B.-Q.,  2021, arXiv e-prints, \href
  {https://ui.adsabs.harvard.edu/abs/2021arXiv211011595Z} {p. arXiv:2110.11595}

\bibitem[\protect\citeauthoryear{{Zucker}, {Battersby}  \& {Goodman}}{{Zucker}
  et~al.}{2015}]{2015ApJ...815...23Z}
{Zucker} C.,  {Battersby} C.,   {Goodman} A.,  2015, \mn@doi [\apj]
  {10.1088/0004-637X/815/1/23}, \href
  {http://adsabs.harvard.edu/abs/2015ApJ...815...23Z} {815, 23}

\makeatother
\end{thebibliography}

\clearpage
\appendix
\section*{Deriving 3D velocities}

Our sample is taken from our previous paper \citep{2021arXiv211011595Z}, which contains a sample of YSO
associations within $\sim 3 \rm kpc$ from the Sun. We identify their molecular cloud counterparts from the CO observations \citep{2001ApJ...547..792D} by visual inspection. The radial velocities of the clouds are calculated by fitting Gaussians to the line profiles of the individual subfields overlapped to the cloud. 


The radial velocities are combined with the Gaia data release 2 (Gaia DR2) \citep{ 2018a&a...616a...1g} measurements of proper motions and parallaxes of the YSO associations to derive 3D velocities. 
The Gaia measurements are represented in the Barycentric Reference
Frame, whose center locates at the center of gravity of the Solar system. 
We have obtained the transverse velocities of the individual YSO associations from their Gaia 
DR2 proper motions and parallaxes \citep{2021arXiv211011595Z}. The radial velocities are obtained from a CO survey \citep{2001ApJ...547..792D}, where the majority of the Milky Way disk is covered. For each complex, the transverse velocities can be measured to the accuracy of $\lesssim 1 \;\rm km\;s^{-1}$. Limited by the velocity resolution of CO observations, the radial velocity measurements have uncertainties of around 1.3 $\rm km\;s^{-1}$. The combined uncertainties are within 2 $\rm km\;s^{-1}$, which is small.
In radio observations, the radial velocities are already converted into the Local Standard of Rest (LSR) by assuming $(u_{\odot}^{\rm std} = 10.3\; {\rm km\;s^{-1}}, v_{\odot}^{\rm std}=15.3\; {\rm km\;s^{-1}}, w_{\odot}^{\rm std}=7.7\; {\rm km\;s^{-1}} )$, which values are outdated \citep{2009ApJ...700..137R}. Radial velocities derived in this fashion are called the $v_{\rm LSR, std}$. Our first step is to convert the $v_{\rm LSR, std}$ back to $v_{\rm
Barycentric}$.

We then combine the radial and transverse velocity measurements to derive the 3D velocities in the Barycentric frame,
obtaining $\vec{v}_{\rm Barycentric}$ and the corresponding locations
$\vec{r}_{\rm Barycentric}$. In the next step, we correct for the effect that the Sun
is 20.8 \,pc above the Galactic Plane \citep{2019MNRAS.482.1417B}, and is 8.34 \, kpc from the Galaxy center \cite, and drive the 3D locations of the YSO
associations in the kinematic LSR reference frame $\vec{x}_{\rm k-LSR}$. The velocities in the kinematic LSR reference  frame $\vec{v}_{\rm k-LSR}$  are obtained assuming $(u_{\odot} = 11.69 \;{\rm km\;s^{-1}}, v_{\odot}=10.16 \;{\rm km\;s^{-1}}, w_{\odot}=7.6\; {\rm km\;s^{-1}} )$ \citep{2021MNRAS.504..199W}.

\section*{Reference frames}

We further project the 3D LSR velocities of the individual YSO-MC complexes into different frames for our convenience. 
We adopt the dynamical-LSR, and define new frames called the  de-accelerated-LSR frame and define the Local Co-rotating Frame (LCF). In the de-accelerated-LSR, a net acceleration caused by the circular motion of the sun, which is of order $v_{\rm circ}^2/R_0$, is removed. In the LCF, a net rotation of order $v_{\rm circ}/ R_0$ is further removed to reveal shear. 
Finally, Galactic shear can be removed to reveal the peculiar velocities of the clouds, where a rotation curve shall be assumed. The difference between these reference frames is summarized in Table \ref{tab:frames}, where  $R_0$ is the Sun's Galactocentric distance, and $v_{\rm circ}$ is the velocity of a  perfectly circular orbit around the center of the galaxy at the Sun's galactocentric distance.

\begin{table*}
    \centering
    \begin{tabular}{|| c c c c ||} 
     \hline
       &  LSR&  de-accelerated LSR & Local Co-rotating Frame (LCF)  \\
     \hline
     Velocity of the Center & $(0, v_{\rm circ}, 0)$ &   $(0, v_{\rm circ}, 0)$  &   $(0, v_{\rm circ}, 0)$ \\
     \hline
     Acceleration & 0 & $a = v_{\rm circ}^2/ R_{0}$ &$a = v_{\rm circ}^2/ R_{0}$   \\
     \hline
     Frame Rotation & 0 & 0 & (0, 0, -$v_{\rm circ} / R_{0}$)\\
     \hline
   Notes & non-inertial & inertial & similar to shearing-box simulations.   \\
\hline
Centered around &          \multicolumn{3}{c}{The Sun}           \\
   \hline 

    \end{tabular}
    \caption{Different reference frames discussed in the current work.  $R_0$ is the Sun's Galactocentric distance, and $v_{\rm circ}$ is the velocity of a  perfectly circular orbit around the center of the galaxy at the Sun's galactocentric distance. \label{tab:frames}}
    \label{table:1}
    \end{table*}

    The Kinematic LSR is a non-rotating frame that follows the mean motion of the stars in the Solar vicinity. The frame itself has no acceleration. This frame is non-inertial, as it experiences
    an acceleration from the center of the Galaxy, e.g. $ |\vec{a}| = v_{\rm circ}^2 / R_0$, where  $v_{\rm circ}$ is the rotation speed which should be considered when modeling the dynamics, and $R$ is the Galactocentric distance in the Galactic plane. 

    The dynamical-LSR also follows the mean motion of stars. Compared to the kinematic-LSR, this frame follows a circular orbit, where the acceleration towards the Galaxy center is removed. The dynamical-LSR is an inertial frame. 

    The local co-rotating frame is a frame defined in this paper. Similar to the dynamical-LSR, the frame follows a circular orbit around the center of the Galaxy. In addition, it contains a rotation around the $Z$ axis. The rotational speed is of order $v_{\rm circ} / R_{0}$. We added this rotation to ensure that the $X$-axis of the frame is locked toward the center of the Galaxy. In this frame, we can observe the motion of gas in a frame similar to those adopted in the shearing-box simulations of accretion disks.
    Because of the added rotation, the frame is non-inertial.  Calculations done in this frame should take the Coriolis force into account.

\begin{figure*}
    \includegraphics[width = 0.9\textwidth]{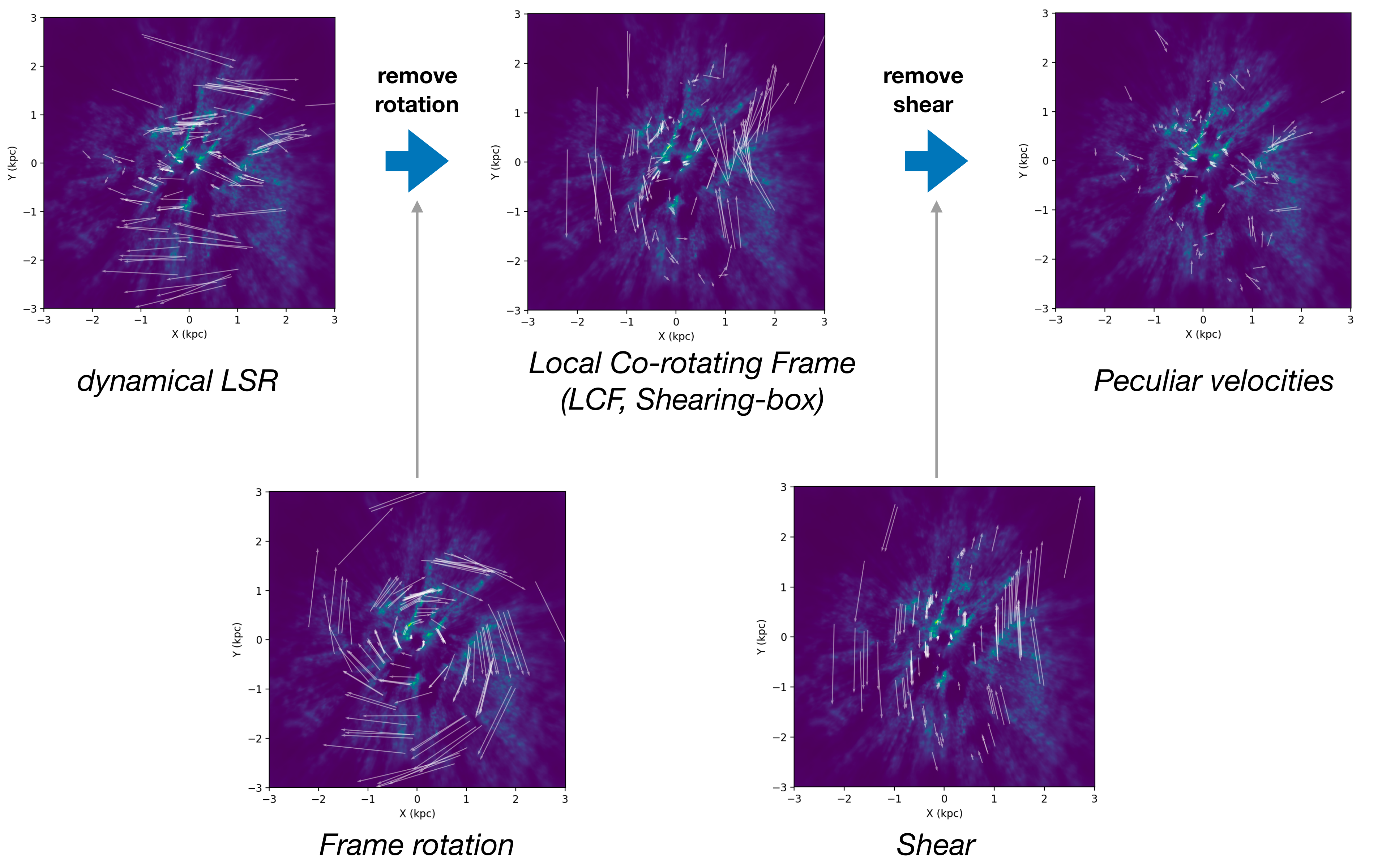}
    \caption{Velocity field viewed in the dynamical LSR frame, Local Co-rotating Frame, a plot of the peculiar velocities, as well as the the conversions between them. \label{fig:conversion} }
\end{figure*}

    Differences between different reference frames as well as the conversions are illustrated in Fig. \ref{fig:conversion}. Finally, to reveal the peculiar motions, we further
    subtract the expected shear assuming Keplerian rotations, and derive $v_{\rm peculiar}$. In the solar vicinity, at location $\vec{x}$, the rotation curve is approximated as 
   $v_{\rm circ, \vec{x}} = (R_{\vec x} - R_0) (A + B) + v_{\rm circ}$, where $A
   $ and $B$ are Oort constants \citep{1927BAN.....3..275O} with updated values \citep{2021MNRAS.504..199W}. 

\section*{Forecasting Cloud Evolution}
\begin{figure}
    \includegraphics[width = 0.5\textwidth]{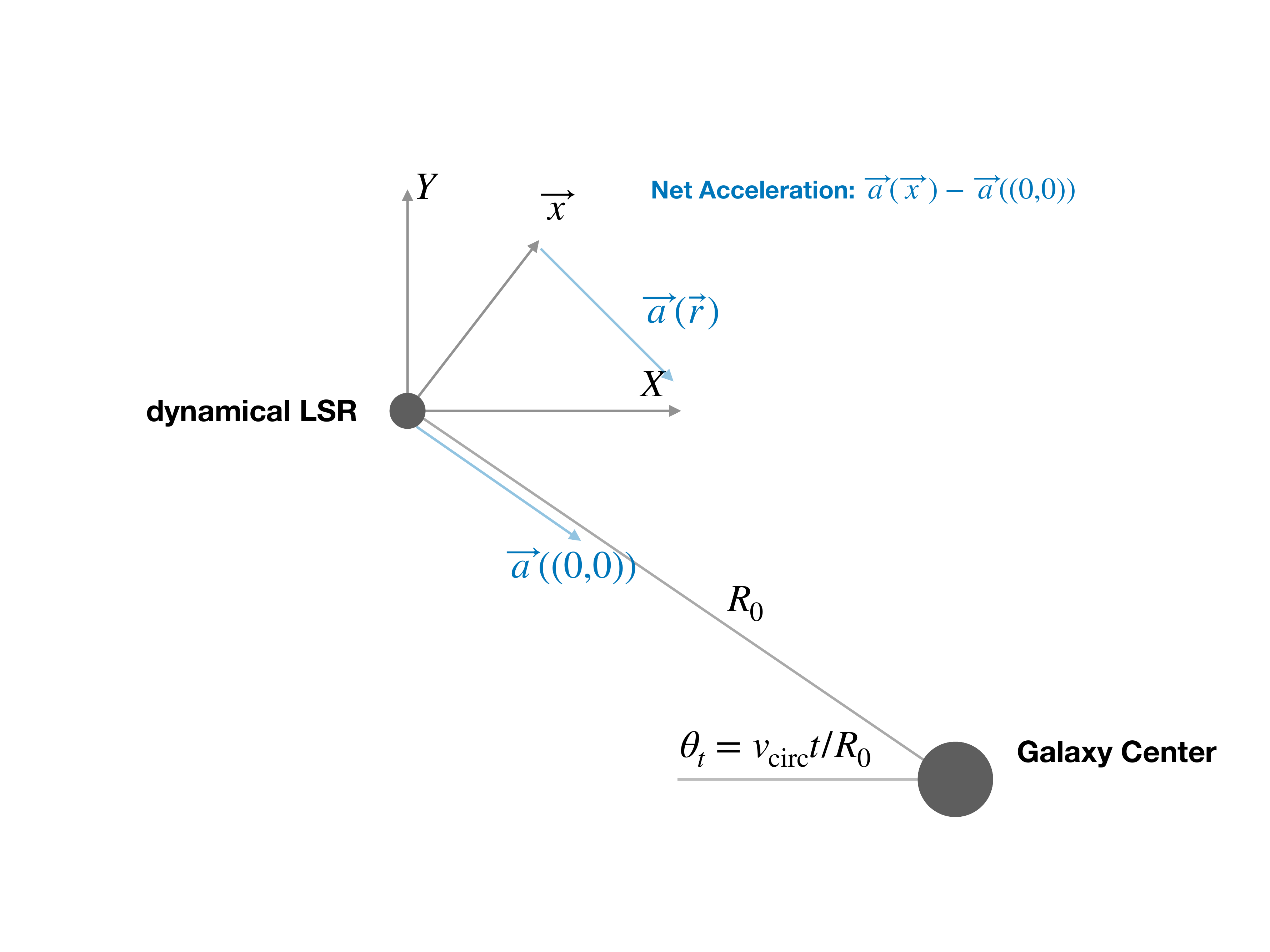}
    \caption{\label{fig:configuration} The  configuration used in our simulation
    forecast. In the dynamical LSR frame, a moving body experience experiences a
    residual acceleration should be computed as $\vec{a}({\vec{r}})_{\rm
    dynamical-LSR} = \vec{a}(\vec{r}) - \vec{a}((0, 0))$, where $(0,0)$ is the
    origin.  } 
\end{figure}

We compute the motion of clouds in the dynamical-LSR reference frame, by solving the following equations
\begin{eqnarray}
    \dot{ \vec{x}}(t) &=&  \vec{v} (t)\\ \nonumber
    \dot{ \vec{v}} (t) &=& \vec{a}(\vec{r})\;,
\end{eqnarray}
where the computation is limited to the $X$-$Y$ plane, e.g. $\vec{r} = (X, Y)$ and $\vec{v} = (v_x, v_y)$, where $\vec{r}$, $\vec{v}$ and $\vec{a}$ stand for the positions, velocities and acceleration, respectively. Because this frame rotates around the Galaxy center in a circular orbit, at time $t$, the Galactic center locates at $(R{\rm cos}(\theta_t),  - R {\rm sin}(\theta_t))$ where  $\theta_t = v_{\rm circ} t / R $. Because in the dynamical-LSR, the gravitational force of the Galaxy is cancelled out by the circular motion, at $(0, 0)$ the acceleration is zero, and the residual acceleration at location $\vec{x}$ is 
\begin{equation}
    \vec{a}({\vec{r}})_{\rm dynamical-LSR} = \vec{a}(\vec{r}) - \vec{a}((0, 0)) \;,
\end{equation}
where 
\begin{eqnarray}
    \vec{a}(\vec{r}) &=& (a_x, a_y) \\ \nonumber
    a_x &=& a_{0} \;{\rm cos} \theta \\\nonumber
    a_y &=& - a_{0}\; {\rm sin} \theta\\\nonumber
    \theta&=& {\rm arctan}(\frac{R {\rm sin} \theta_t}{R {\rm cos\theta_t} - X  }) \\ \nonumber
    a_{0} &=& \frac{v_{\rm circ, \vec{r}}^2}{R_{\vec{r}}}
\end{eqnarray}
and 

\begin{eqnarray}
    \vec{a}((0,0 )) &=& (a_x, a_y) \\ \nonumber
    a_x &=& a_{0} \;{\rm cos} \theta_x \\\nonumber
    a_y &=& - a_{0}\; {\rm sin} \theta_y\\\nonumber
    \theta &=& \theta_t \\ \nonumber
    a_{0x} &=& \frac{v_{\rm circ}^2}{R_0}
\end{eqnarray}
where $v_{\rm circ}$ is the circular velocity measured at the origin, $R_{0}$ the Galactocentric distance of the origin, $v_{\rm circ, \vec{r}}$ the circular velocity measured at the 
location $\vec{r}$, and $R_{\vec{r}}$ is Galactictocentric distance of the position $\vec{r}$. The circular  velocity can be approximated using 
\begin{equation}
    v_{\rm circ, \vec{r}} = (R_{\vec{r}} - R_{0}) (A + B) + v_{\rm circ, 0}
\end{equation}
where $A = 16  {\; \rm km \; s^{-1}\; kpc^{-1} }$ and $B = -12  {\; \rm km\;
s^{-1}\; kpc^{-1}}$ \citep{2021MNRAS.504..199W} are Oort constants \citep{1927BAN.....3..275O}. This frame is chosen only for our
convenience. One can, of course, perform the calculation in a different frame
and obtain identical results. Fig. \ref{fig:configuration} illustrating this
configuration and the acceleration calculation. 

\section*{Measuring bubble expansion}
\begin{figure*}
    \includegraphics[width = 1 \textwidth]{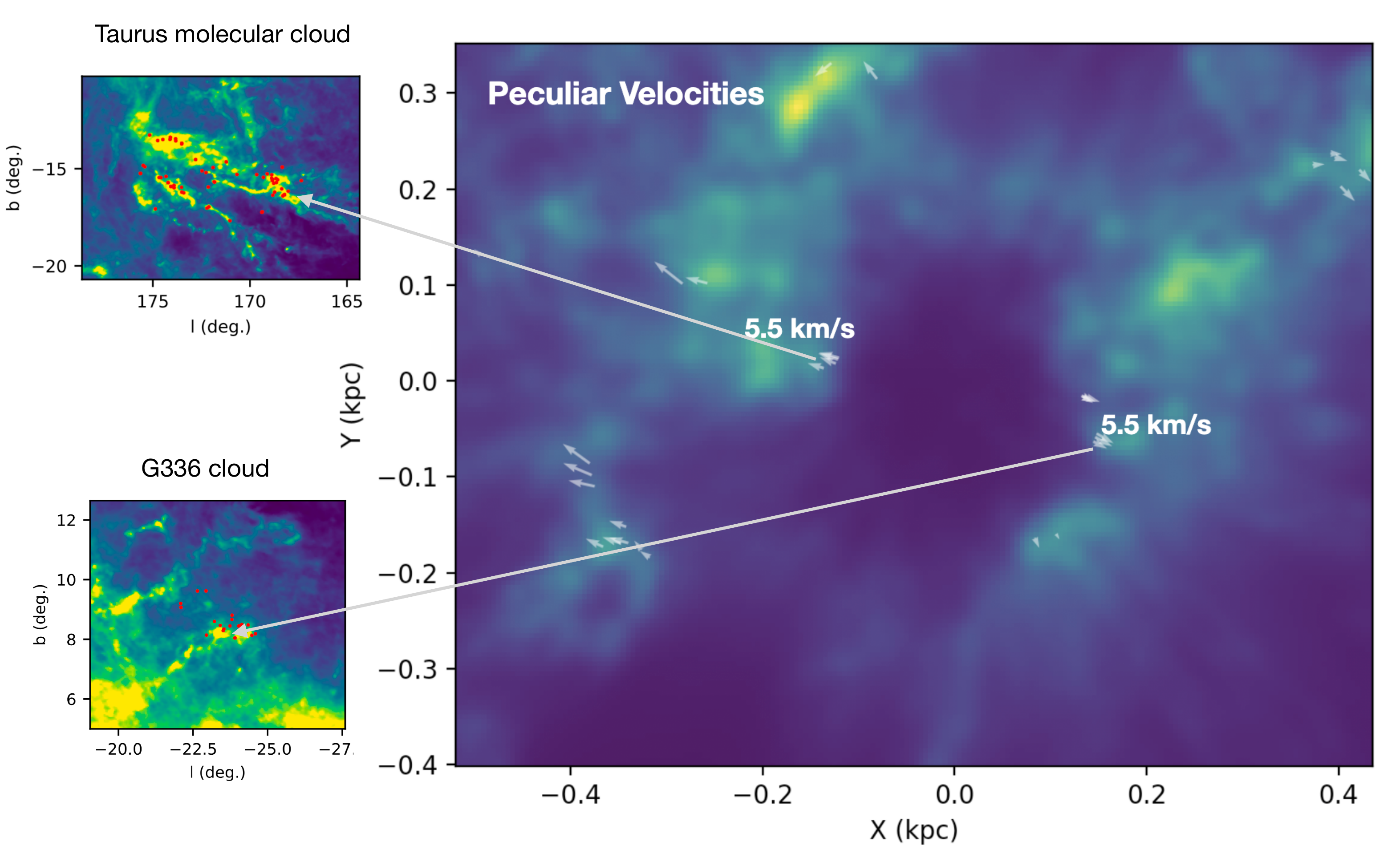}
    \caption{ Measuring bubble-induced peculiar velocities around the Local Bubble. Background images is a dust map in the Galactic $l$-$b$ produced by \citet{2019A&A...625A.135L}, and the velocities are denoted using arrows. We also added Planck 870 $\mu$m images 
    of two the clouds which are used to measure the expansion velocity. In these plots, the locations of the member YSOs are indicated using red dots. \label{fig:expansion}} 
\end{figure*}

We choose two groups of YSO-MC complexes located at the opposites of the Local Bubble to measure the expansion velocity, each group has an expansion-induced peculiar velocity of around 5.5 $\rm km\;s^{-1}$,  they are separated by 0.3 kpc. This is illustrated in Fig. \ref{fig:expansion}.

\clearpage
\onecolumn

\end{document}